\documentstyle[floats,twocolumn,aps,prb] {revtex}

\begin{document}
\draft
\tighten

\title{{\it Ab initio} Quantum and {\it ab initio} Molecular Dynamics\\ 
        of the Dissociative Adsorption of Hydrogen~on~Pd(100)} 

\author{Axel Gross and Matthias Scheffler}

\address{Fritz-Haber-Institut der Max-Planck-Gesellschaft, Faradayweg 4-6, 
D-14195 Berlin-Dahlem, Germany} 

\date{May 2, 1997}

\maketitle


\begin{abstract}

The dissociative adsorption of hydrogen on Pd(100)
has been studied by {\it ab initio} quantum dynamics and 
{\it ab initio} molecular dynamics calculations.
Treating all hydrogen degrees of freedom as dynamical coordinates
implies a high dimensionality and requires statistical averages over 
thousands of trajectories. An efficient and accurate treatment of such 
extensive statistics is achieved in two steps:
In a first step we evaluate the {\it ab initio} potential energy surface (PES)
and determine an analytical representation. Then, in an independent
second step dynamical calculations are performed on 
the analytical representation of the PES. Thus the dissociation
dynamics is investigated without any crucial assumption
except for the Born-Oppenheimer approximation which is anyhow employed
when density-functional theory calculations are performed. 
The {\em ab initio} molecular dynamics is compared to
detailed  quantum dynamical calculations on exactly the
same {\em ab initio} PES. The occurence of quantum oscillations
in the sticking probability as a function of kinetic energy 
is addressed. They turn out to be very sensitive to the symmetry 
of the initial conditions. At low kinetic energies sticking 
is dominated by the steering effect which is illustrated using 
classical trajectories. The steering effects depends on the kinetic
energy, but not on the mass of the molecules. Zero-point effects lead
to strong differences between quantum and classical calculations of
the sticking probability. The dependence of the sticking probability on 
the angle of incidence is analysed; it is found to be in good agreement 
with experimental data. The results show that the determination of the 
potential energy surface combined with high-dimensional dynamical 
calculations, in which all relevant degrees of freedon are taken into 
account, leads to a detailed understanding of the dissociation dynamics 
of hydrogen at a transition metal surface.
\end{abstract}



\section{Introduction}

The dissociative adsorption of molecules on surfaces is one of the
fundamental reaction steps occuring in catalysis, corrosion, 
and in the hydrogen gas storage in metals.
Adsorption corresponds to a process in which statistically distributed
molecules hit the surface from the gas phase. For a diatomic molecule
this requires the calculation of thousands of trajectories; 
the adsorption probability is then obtained by averaging over
these trajectories. In ``traditional'' {\it ab initio} molecular dynamics
the electronic structure, total energy and the forces acting on the
nuclei are determined for each configuration {\em during the journey}
of the particles which requires of the order of 100 -- 1000 self-consistent
calculations for each trajectory. Since the determination of total
energies and forces is still a heavy computational task, the number
of trajectories obtainable in such a ``during the journey'' {\it ab initio}
molecular dynamics simulation is limited to numbers well 
below 100.\cite{DeVita93,Gro97H2Si} These numbers are too low
to achieve a reasonable statistics. Typically one needs to consider
of the order of 10$^3$ to 10$^5$ trajectories, 
depending on the kind of experiment to be simulated.\cite{Gro97Vac} 
Therefore we propose a ``divide and conquer'' approach for {\it ab initio} 
molecular dynamics: At first the road map should be created and only
then the journey started. Thus, in a first (elaborate) study we analyse
the {\it ab initio} potential energy surface (PES). This PES is 
parametrized in a suitable form and only then the molecular dynamics 
calculations are performed on the analytical representation of the 
{\it ab initio} potential. In this way it is easy to study
up to 10$^6$ trajectories.

\vspace{-12.6cm}

\hfill {\tt subm.~to~Phys.~Rev.~B}

\vspace{12.25cm}

If hydrogen is involved in the dissociation process, quantum effects 
may play a role in the dynamics.
We have recently improved a quantum dynamical algorithm that makes it
possible to treat {\em all} six degrees of freedom of the hydrogen
molecule in the dissociation process quantum dynamically.\cite{Gro95a}
Quantum effects can thus be determined by comparing
the results of the quantum dynamical calculations
with classical trajectory calculation on exactly the same PES.
It turns out that for hydrogen moving on a strongly corrugated and
anisotropic PES zero-point effects can be substantial.\cite{Gro97Vac} 
One advantage of the quantum dynamics, that to our knowledge has not been 
emphasized in the literature yet, is that the averaging over the 
initial conditions is done
automatically. For example, a plane wave in a $j=0$ rotational state
describing the incident molecular beam hits the surface everywhere 
in the surface unit cell and contains all molecular orientations with 
equal probability.

Dissociative adsorption systems can be roughly divided into two 
classes \cite{Ren94,Hol94,Dar95,Gro95JCP,Gro96SSa}: 
Systems, where the sticking probability is monotonously increasing 
as a function of the incident kinetic energy of the impinging
molecules, and systems, where the sticking probability shows an
initial decrease with increasing kinetic energy.
The in most detail studied system H$_2$/Cu 
\cite{Hay89,Ang89,Hod92,Ret92,Ret95,Kue91,Dar92a,Mow93,Dar94,Bru94,
Gro94,Kum94,Dai95,Kro96,Din97} belongs to the first class. These systems
are characterized by a  minimum barrier hindering dissociative
adsorption, so that increasing the kinetic energy helps to overcome
the barrier.

The second class consists of adsorption systems like  
H$_2$\cite{Ren89,Res94,Ber92,But94,Aln89,But95,Dix94}, 
O$_2$\cite{Lun88,Bra96,Ret86,Whe96}, and N$_2$\cite{Ret88} 
on various transition metal surfaces. In particular the 
well-studied system H$_2$/Pd(100),\cite{Ren94,Dar95,Ren89,Sch92,Dar92,Bre94}
 which is the subject of our study, belongs to this class.  
An initially decreasing sticking probability had usually been explained 
by a {\em precursor mechanism}.\cite{Ren94} 
In this concept the molecules are temporarily trapped in a 
molecular physisorption state, the so-called {\em precursor state}, 
before they dissociatively adsorb. 
The energy dependence of the sticking probability is related to the 
trapping probability into the precursor state.
It is this trapping probability which decreases with increasing 
energy.\cite{Ren94}

However, it has for example been shown for the system 
H$_2$/W(100)--c(2$\times$2)Cu \cite{But95} that for a hydrogen molecule 
impinging on a metal substrate the energy transfer to substrate phonons 
is much too small to account for the high sticking 
probabilities at low kinetic energies due to the large mass mismatch. 
Therefore also direct non-activated 
adsorption together with a steering effect has been suggested in order to 
explain the initial decrease of the sticking probability by King almost twenty
years ago \cite{King78}. Still, in low-dimensional dynamical treatments of the
H$_2$/Pd(100) system no steering effect was observed.\cite{Sch92,Dar92,Bre94}
Only very recently it has been shown by high-dimensional
quantum dynamical calculations based on {\it ab initio} potential energy 
surfaces (PES) for the systems H$_2$/Pd(100) \cite{Gro95a,Gro95JCP,Wil95} 
and H$_2$/W(100) \cite{Kay95,Whi96} that indeed an initial decrease of the
sticking probability with increasing kinetic energy is not necessarily
due to a precursor mechanism. For both systems the PES has non-activated
paths towards dissociative adsorption and no molecular
adsorption well. However, the majority of pathways towards dissociative 
adsorption has in fact energy barriers with a rather broad
distribution of heights and positions, i.e. the PES is strongly anisotropic 
and corrugated. Similiar features of the potential have recently been found 
also for the interaction of H$_2$ with Rh(100).\cite{Eich96} A slow molecule 
moving on such a PES with an unfavorable initial configuration for
dissociative adsorption can be steered efficiently towards non-activated
paths to adsorption by the forces acting upon the molecule. This mechanism
becomes less efficient at higher kinetic energies because then the molecule
is too fast to be diverted significantly. More particles are therefore 
scattered back into the gas-phase from the repulsive part of the potential.
This leads to the initial decrease 
of the sticking probability. At even higher kinetic energies the molecule
will eventually have enough energy to directly traverse the barriers
causing the sticking probability to rise again.

In our calculations for the interaction of H$_2$/Pd(100)
{\em all} six degrees of freedom of the hydrogen molecule are treated
dynamically \cite{Gro95a}. This makes it possible to investigate the
influence of {\em all} hydrogen degrees of freedom on the dissociative
adsorption, scattering and associative desorption on an equal footing.
So far we have studied the dependence of adsorption
and desorption on kinetic energy, molecular rotation and orientation
\cite{Gro95a,Gro96SSb}, molecular vibration \cite{Gro96CPLa}, ro-vibrational
coupling \cite{Gro96Prog} and the rotationally elastic and inelastic 
diffraction of H$_2$/Pd(100) \cite{Gro96CPLb}.

In this article we will first describe the quantum 
and classical methods we have used to determine the adsorption dynamics
of hydrogen on Pd(100). Then we address the origin of oscillations
in the sticking probability as a function of the kinetic energy.
Next the steering effect is illustrated and the differences between 
classical and quantum dynamics are discussed as are isotope effects in
the dissociative adsorption dynamics. Finally we will focus on the
dependence of the sticking probability on the angle of incidence.

\section{Computational details}

In our approach the dynamical simulations including all relevant degrees
of freeedom are performed on an analytical representation of the
{\it ab initio} PES. Thus in principle we apply only one approximation,
namely the Born-Oppenheimer approximation, i.e., we assume that the
electrons follow the motion of the nuclei adiabatically. 
Obviously in practice there is a second important approximation,
namely the treatment of the exchange and correlations effects in
the density functional theory calculations.

As far as the number of relevant degrees of freedom is concerned,
in the case of hydrogen dissociation on densely packed metal surfaces
usually no significant surface rearrangement upon adsorption occurs, and 
there is only a small energy transfer from the light hydrogen molecule to 
the heavy substrate atoms. Even if there is any surface relaxation upon
hydrogen adsorption, it occurs typically on a much larger time scale
than the adsorption event. The crucial process in the dissociative 
adsorption for these particular systems is therefore the conversion of 
translational and internal energy of the hydrogen molecule into 
translational and vibrational energy of the adsorbed hydrogen atoms. 
Thus the dissociation dynamics can be described by a six-dimensional PES
which takes only the molecular degrees of freedom into account.
In the following we present our formalism in such a six-dimensional
formulation. In principle, however, it can be extended to
include also the substrate degrees of freedom if they are relevant.

\subsection{Parametrization of the {\it ab initio} potential}
\label{Para}

\begin{figure}[tb]
\unitlength1cm
\begin{center}
   \begin{picture}(10,9.0)
      \includegraphics{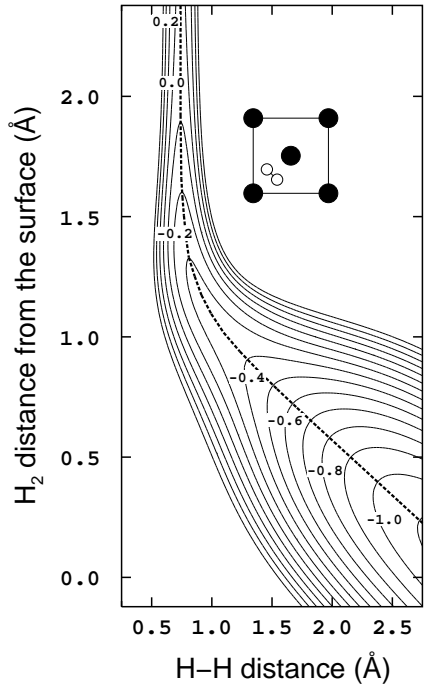}
   \end{picture}

\end{center}
   \caption{Contour plot of the PES along a
            two-dimensional cut through the
            six-dimensional coordinate space of H$_2$/Pd\,(100).
            The inset shows the
            orientation of the molecular axis and the lateral
            H$_2$ center-of-mass coordinates, i.e. the coordinates
            $X$, $Y$, $\theta$, and $\phi$. The coordinates 
            in the figure are the H$_2$ center-of-mass distance 
            from the surface $Z$ and the H-H interatomic distance $r$. The 
            dashed line is the optimum reaction path.
            Energies are in eV per H$_2$ molecule.
            The contour spacing is 0.1~eV.  }

\label{elbow}
\end{figure}

In order to obtain a reasonable analytical representation of the PES,
first a sufficient number of {\it ab initio} total energies 
has to be computed. High-symmetry points of the multi-dimensional 
configuration space are reflected by extrema in the PES. 
Typically in between the high-symmetry points the PES is smooth and has no
additional maxima and minima. Of course, this assumption has to
checked carefully. In the case of a rigid surface, the PES of a
diatomic molecule interacting with this surface is a function of
the six molecular degrees of freedom. They can be represented, e.g.,
by the center-of-mass coordinates $X,Y,Z$\,, the interatomic distance $r$,
and the polar and azimuthal angle of the molecular axis $\theta$ and
$\phi$. Figure~\ref{elbow} shows a two-dimensional cut through the 
six-dimensional coordinate space of H$_2$/Pd\,(100),
a so-called elbow plot. The two
considered coordinates are the H$_2$ center-of-mass distance
from the surface $Z$ and the H-H interatomic distance $r$.

In order to solve the time-independent Schr{\"o}dinger equation
describing dissociative adsorption it is advantageous  to transform
the coordinates in the $Zr$-plane into reaction path coordinates
$s$ and $\rho$ \cite{Hof63,Mar64,Bre89}. 
Here $s$ describes the position along the ``reaction
path'' -- the dashed line in Fig.~\ref{elbow} -- and $\rho$ is the
coordinate perpendicular to $s$ (see section \ref{Qdyn}).
We have then parametrized the function $ V (X, Y, s, \rho, \theta, \phi)$,
which  describes the potential energy surface on which the hydrogen molecule
moves, in the following form:\cite{Gro95a}
\begin{equation}
V \ (X,Y,s,\rho,\theta,\phi) \ = \ V^{\rm corr} \ + \ V^{\rm rot} \ 
+ \ V^{\rm vib}
\end{equation}
with
\begin{equation}
V^{\rm corr} \ = \ \sum_{m,n = 0 }^2 \ V_{m,n}^{(1)} (s) \ \cos mGX \ \cos nGY,
\end{equation}
\begin{eqnarray}
\lefteqn{V^{\rm rot} = \sum_{m=0}^1 \ V_m^{(2)}(s) \ \frac{1}{2} 
\cos^2 \theta \  (\cos mGX + \cos mGY)} \nonumber\\
& & + \sum_{n=1}^2 \ V_n^{(3)} (s) \ \frac{1}{2} 
 \sin^2 \theta \ \cos 2 \phi \ (\cos nGX - \cos nGY) 
\label{Vrot}
\end{eqnarray}
and
\begin{equation}
V^{\rm vib} \ = \ \frac{\mu}{2}\ \omega^2 (s)\ [\rho \ 
- \ \Delta \rho (X,Y,s)]^2 \ .
\label{V_vib}
\end{equation}
$G = 2 \pi / a$ is the length of the basis vectors of the square surface 
reciprocal lattice, $a$ is the nearest neighbor distant between  Pd atoms
and $\omega (s)$ is the vibrational frequency. We note that
Wiesenekker {\it et al.} \cite{Wie96} have recently employed an equivalent
analytic representation to describe the 6D-PES of H$_2$/Cu(100), 
the only difference being that they use cartesian coordinates in the 
$Zr$-plane instead of reaction path coordinates.

It turns out that the calculation of total energies for 250 different 
configurations is sufficient to determine the parameters
necessary to describe the functions appearing in the potential
parametrization. The {\it ab initio} energies are obtained\cite{Wil95}
using density functional theory together with the generalized gradient 
approximation (GGA) \cite{Per92} for the exchange and correlation
functinal and the full-potential linear augmented plane wave method
(FP-LAPW) (see Ref.~\onlinecite{Koh96} and references therein).
The ten functions $V_{m,n}^{(i)}(s)$ and $\omega (s)$
are determined such that the difference to the 
{\it ab initio} calculations on the average is smaller than 25~meV.

The introduction of the displacement $\Delta \rho$ in the potential
term $V^{\rm vib}$ (Eq.~\ref{V_vib}) takes into account that
the location of the minimum energy path in the $Zr$-plane depends
on the cut through the six-dimensional configuration space
of H$_2$/Pd(100). $\Delta \rho$ does not influence the barrier
distribution, however, it changes the curvature of the minimum
energy paths in the $Zr$-planes. For the calculated elbow plots
$\Delta \rho$ reaches values of up to 0.6~{\AA}, but only for large
separations of the hydrogen atoms, i.e., when the molecule is
already dissociated. Large values of $\Delta \rho$ require a
large number of vibrational eigenfunctions in the expansion of
the hydrogen wave function in the coupled channel scheme (see below),
which makes the calculations very time-consuming. Since the
large values of $\Delta \rho$ occur only for large separations of
the hydrogen atoms, they do not influence the calculated sticking
probabilities and scattering properties significantly, as we have
checked by test calculations. 
We have therefore parametrized the displacement
properly only for values of $|\Delta \rho| \le 0.15$~{\AA}.

\subsection{Quantum dynamics}
\label{Qdyn}


The Hamiltonian of a hydrogen molecule interacting with a rigid surface
can be written as
\begin{equation}
H \ = \ -\frac{\hbar^2}{2M} \ \nabla_{\bf R}^2 \ - 
\ \frac{\hbar^2}{2\mu} \  \nabla_{\bf r}^2 \ + 
\ V ({\bf R}, {\bf r}) \ .
\end{equation}
${\bf R} = ({\bf r}_1 + {\bf r}_2)/2 = (X, Y, Z)$ 
and ${\bf r} = ({\bf r}_2 - {\bf r}_1)$
are the center-of-mass and relative coordinates of the hydrogen molecule,
respectively, $M = 2m$ and $\mu = m/2$ are the total and reduced mass of
the hydrogen molecule, respectively, where $m$ is the mass of a hydrogen
atom. Now we write the relative part in spherical coordinates; 
with $r = |{\bf r}|$ this yields
\begin{eqnarray}
H & = & -\frac{\hbar^2}{2M} \ \nabla_{\bf R}^2 \ - \
\frac{\hbar^2}{2\mu} \ \left( \frac{\partial^2}{\partial r^2} \
+ \ \frac{2}{r}\frac{\partial}{\partial r} \ 
- \ \frac{{\bf L}^2}{r^2} \right) \nonumber\\
& & + \ V ({\bf R}, r, \theta, \phi) \ ,
\end{eqnarray}
where ${\bf L}$ is the angular momentum operator. 

We define $r_e$ as
the equilibrium bond length for a certain molecular configuration, i.e.,
$r_e$ is in general a function of the five coordinates $X, Y, Z, \theta$ and
$\phi$. The vibrational amplitude of diatomic molecules is usually
small compared to the equilibrium bond length, which means that
\begin{equation}
| r_e - r | \ \ll \ r \ .
\end{equation}
This allows us \cite{Haken} to neglect the term 
$\frac{2}{r}\frac{\partial}{\partial r}$ in the Hamiltonian and to
approximate the angular momentum term by
$\frac{{\bf L}^2}{r^2} \approx \frac{{\bf L}^2}{r_e^2}$.
Then we end up with the following Hamiltonian
\begin{eqnarray}
\tilde H & = & -\frac{\hbar^2}{2M} \ \nabla_{\bf R}^2 \ - \
\frac{\hbar^2}{2\mu} \ \left( \frac{\partial^2}{\partial r^2} \
- \ \frac{{\bf L}^2}{r_e^2} \right) \nonumber\\
& & + \ V ({\bf R}, r, \theta, \phi) \ ,
\end{eqnarray}

As already mentioned in section \ref{Para}, it is advantageous  to transform
the coordinates in the $Zr$-plane into reaction path coordinates
$s$ and $\rho$ in order to solve the time-independent Schr{\"o}dinger equation
describing dissociative adsorption. To perform the transformation
to the reaction path coordinates, first $Z$ has to be
mass-scaled to the coordinate $z$ by
\begin{equation}
z \ = \ Z \ \sqrt{\frac{M}{\mu}} \ .
\end{equation}
The PES in Fig.~\ref{elbow} is already plotted 
according to mass-scaled coordinates.
With the reaction path coordinates $s$ and $\rho$ the Hamiltonian
becomes
\begin{eqnarray}
\tilde H & = & -\frac{\hbar^2}{2\mu} \ \left( 
\eta^{-1} \frac{\partial}{\partial s} 
\left( \eta^{-1} \frac{\partial}{\partial s} \right) \ 
+ \eta^{-1} \frac{\partial}{\partial \rho}
\left( \eta \frac{\partial}{\partial \rho} \right)\right) \nonumber\\
& & -\frac{\hbar^2}{2\mu} \ \left( 
\frac{\partial^2}{\partial r^2} \ 
- \ \frac{{\bf L}^2}{r_e^2} \right) \nonumber\\
& & -\frac{\hbar^2}{2M} \ 
\left( \frac{\partial^2}{\partial X^2} + 
\frac{\partial^2}{\partial Y^2} \right)
+ \ V (X, Y, s, \rho, \theta, \phi) \ . \label{ReacHam}
\end{eqnarray}
The coupling parameter $\eta$ is defined by
\begin{equation}
\eta \ = \ 1 \ - \kappa(s)\ \rho \ ,
\end{equation}
where $\kappa(s)$ is the curvature of the lowest energy reaction path
(the dashed line in Fig.~\ref{elbow}).

The displacement $\Delta \rho$ also enters the denominator of the 
angular momentum term in the Hamiltonian via the equilibrium 
bondlength $r_e$
\begin{equation}
r_e \ = \ r_e^0 (s) \ + \ \sin \phi_r (s) \ \Delta \rho (X,Y,s) \ .
\end{equation}
Here $\sin \phi_r (s)$ is the angle between the reaction path
(dashed line in Fig.~~\ref{elbow}) and the $z$-axis.
Since the relation
\begin{equation}
\sin \phi_r (s) \ \Delta \rho (X,Y,s) \ \ll \  r_e^0 (s) 
\end{equation}
holds, we have expanded $r_e^{-2}$ in the angular momentum term
in a Taylor series in $\sin \phi_r (s) \ \Delta \rho (X,Y,s)$ 
up to second order.

The quantum dynamical calculations are performed by solving the 
time-independent
Schr\"odinger equation for the two hydrogen nuclei moving on the 
six-dimensional PES in a coupled-channel scheme. As channels the 
eigenfunctions of the Hamiltonian for molecules in the gas phase 
are used. We use the concept of the {\em local reflection matrix} 
(LORE)\cite{Bre94,Bre93}. For 
a detailed description of this stable coupled channel method
we refer to Refs.~\onlinecite{Bre94} and \onlinecite{Bre93}. 
In the LORE scheme the reflection matrix $R$ is determined; 
in order to obtain sticking probabilities $S_i$ for some initial 
state $i$, where $i$ stands for a multi-index, we use unitarity:
\begin{equation} 
S_i \ = \ 1 \ - \ \sum_j |R_{ji}|^2 \ .
\end{equation}
$R_{ji}$ is the differential reflection amplitude;
the sum over $j$ extends over all possible reflection states.

The basis set used in the coupled-channel algorithm for the H$_2$ results
presented here included rotational eigenfunctions with rotational 
quantum numbers up to $j_{\rm max} = 8$, vibrational eigenfunctions 
with vibrational quantum numbers up to $v_{\rm max} =2$, and parallel 
momentum states with maximum parallel momentum $p_{\rm max} = 7 \hbar G$ 
with $G= 2\pi/a$. The convergence of the results with
respect to the basis set has been carefully checked by calculations
with maximum quantum numbers $j_{\rm max} =10$, $v_{\rm max} = 3$, and 
$p_{\rm max} = 10 \hbar G$, respectively.

Due to the higher mass of D$_2$ the energy spacing between the quantum levels
is smaller for D$_2$ than for H$_2$. Therefore much more eigenfunctions
in the expansion of the wavefunction have to be taken into account in
the coupled-channel calculations for D$_2$ than for H$_2$. This makes
a six-dimensional quantum treatment of D$_2$ very time-consuming. In order 
to investigate isotope effects we have therefore performed five-dimensional
vibrationally adiabatic quantum calculations for D$_2$, where the molecules
are not allowed to make vibrational transitions. We have already
shown that vibrationally adiabatic calculations are very close to
the full six-dimensional results for the dissociation
of H$_2$ on Pd(100)~\cite{Gro96CPLa}. This should also be valid
for D$_2$ since the ratio of the vibrational time-scale to the
other time-scales of rotation and translation is the same for
H$_2$ and D$_2$. The five-dimensional
quantum calculations for D$_2$ have been performed with rotational
quantum states up to $j_{\rm max} = 12$ and parallel
momentum states with $p_{\rm max} = 11 \hbar G$.

In coupled-channel calculations always the whole $S$-matrix has to be
computed. This leads to a $N^3$-scaling of the algorithm due to
the matrix operations, where $N$ is the number of channels included in 
the calculation. In order to perform these demanding quantum 
dynamical calculations it is therefore necessary
 to utilize the symmetries of the scattering problem
(see also Ref.~\onlinecite{Kro95}). 
First of all selections rules are important. 
Because of the inversion symmetry of the H$_2$ molecule only 
rotational transition with $\Delta j = even$ are allowed, where
$j$ is the rotational quantum number.
In addition, the analytic representation of PES only contains  
rotational potential terms that cause $\Delta m = even$ transitions
(see the second sum of $V^{rot}$ in Eq.~\ref{Vrot}), where
$m$ is the azimuthal quantum number of the H$_2$ molecule.

Furthermore, we exploit the symmetry group of the Hamiltonian
which corresponds to the $C_{4v}$ symmetry of the fcc (100)-surface.
In general the scattering solutions do
not belong to irreducible representations of the symmetry group.
However, if the scattering solutions of interest can be
decomposed into irreducible representations, the number of relevant
channels per coupled-channel calculation can be significantly reduced. 
This is due to the fact that only channels belonging to the same 
irreducible representation of the symmetry group couple to each 
other, since the Hamiltonian commutes with the symmetry operator.

If, for example, the incident parallel momentum corresponds to a 
reciprocal lattice vector (this includes the zero-vector for normal 
incidence) and the initial rotational quantum number $j$ and the azimuthal
quantum number $m$ are even, the scattering solutions can be broken
up into eight different irreducible representations of the symmetry
group, four of which can be further decomposed. In each 
decomposition the number of channels is roughly halved, and in each
irreducible representation the $S$-matrix is calculated separately. 
This leads to a reduction of the computational cost to approximately 
$4\cdot(1/8)^3 + 8\cdot(1/16)^3 = 5/512 \approx 1.0\%$. 
If only the sticking probability for normal incidence is required,
it is sufficient to calculate only two $S$-matrices, i.e.,
the exploitation of the symmetries causes a reduction in the CPU time
to $\sim2\cdot(1/16)^3 = 1/2048 \approx 0.05\%$. Without the use of
the symmetry the calculations presented here would not be feasible.
Using the selection rules and the decomposition into irreducible
representations up to 25,000 channels per total energy are taken into
account in the quantum dynamical calculations; the actual number of
channels in the single calculations is usually $\lesssim 600$.
For a more detailed description of the construction of symmetry
adapted channels, see Ref.~\onlinecite{Kro95}.

\subsection{Classical dynamics}

The classical trajectory calculations are performed on 
{\em exactly the same} PES as the quantum dynamical calculations.
To derive the classical equations of motion from the reaction
path Hamiltonian Eq.~\ref{ReacHam} we have used\cite{Chi93}
\begin{eqnarray}
-i\hbar \ \partial_s & \equiv & p_s \ ,\nonumber \\
-i\hbar \ \partial_\rho & \equiv & p_\rho \ . 
\end{eqnarray}
The equations of motions are numerically integrated with the
Bulirsch-Stoer method with a variable time-step \cite{NumRec}. 
We required that the energy conservation per molecular dynamics run 
was fulfilled to 0.1~meV. The sticking probability is determined by 
averaging over a sufficient number of trajectories. 
The exact number of trajectories 
to be considered depends on the specific initial conditions and ranges 
between 1,815 and 18,330.

As far as the CPU time requirement is concerned, it is a wide-spread
believe that classical methods are much less time-consuming than
quantum ones. This is certainly true if one compares the computational
cost of one trajectory to a quantum calculation. However, 
since in quantum mechanics the averaging over initial conditions
is done automatically while in classical mechanics one
has to average over many trajectories corresponding to different
initial conditions, for the results presented here the quantum method is even
more time-efficient than the classical calculations, in particular if
one considers the fact, that in a coupled-channel method the
sticking and scattering probabilites of all open channels are
determined simultaneously.

\begin{figure}[tb]
\unitlength1cm
\begin{center}
   \begin{picture}(10,6.5)
   \includegraphics{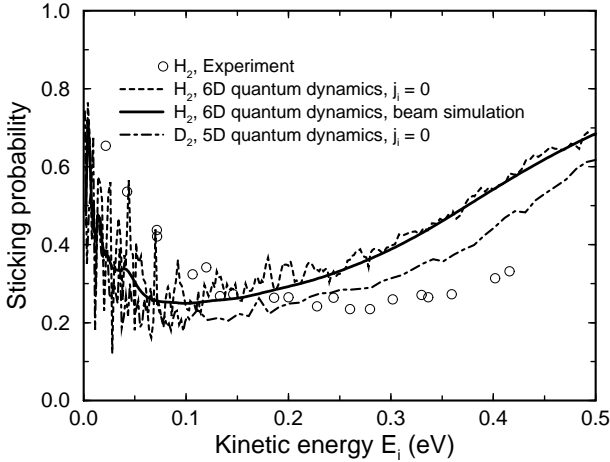}
   \end{picture}
\end{center}
   \caption{Sticking probability versus kinetic energy for
a hydrogen beam under normal incidence on a Pd(100) surface.
Experiment (H$_2$): circles (from ref.~\protect{\onlinecite{Ren89}}); theory:
six-dimensional results for H$_2$ molecules initially in the rotational 
and vibrational ground state (dashed line)
and with an initial rotational and energy distribution 
adequate for molecular beam experiments (solid line), and 
vibrationally adiabatic five-dimensional
results for D$_2$ molecules initially in the rotational ground state 
(dash-dotted line).}
\label{stickiso}
\end{figure}

\section{Results and Discussion}

\subsection{Quantum oscillations}

Figure \ref{stickiso} presents six-dimensional quantum dynamical
calculations of the sticking probability  as a function of the 
kinetic energy of a H$_2$ beam under normal incidence on a Pd(100) surface 
and five-dimensional calculations for D$_2$. In addition, the results 
of the H$_2$ molecular beam experiment by 
Rendulic, Anger and Winkler \cite{Ren89} are shown.

First of all a very strong oscillatory structure is apparent in
the sticking probability as a function of the incident energy.
Such structures reflect the quantum nature of the scattering.
They are known for a long time in He and H$_2$  
scattering \cite{stern} and also in LEED \cite{LEED}.
In the case of the sticking probability of H$_2$/Pd(100), these
oscillations have been the issue of a current debate 
\cite{Gro96CPLb,Ret96PRL,Gro96PRL,Ret96}.
We have recently shown \cite{Gro96CPLb}
that the peaks in the sticking probability can be related to the
opening up of new scattering channels with increasing kinetic energy,
especially at low kinetic energies. 
In particular the emergence of the [10], [11], and [20] diffraction
channels for normal incidence and the opening up of rotational
excitation lead to strong peaks in the sticking probablity.
Here [$n,m$] corresponds to the diffraction indices of the
(100)-surface.

Rettner and Auerbach \cite{Ret96PRL,Ret96} have tried to find
the theoretically predicted oscillations \cite{Gro95a} by
an effusive beam experiment, but they could not detect any. 
As pointed out,\cite{Gro96CPLb,Gro96PRL} the height of the peaks 
is very sensitive to the symmetry of the scattering conditions. 
Surface imperfections
like adatoms and steps and also the thermal motion of the substrate
will reduce the coherence of the scattering process and thus
smooth out the oscillatory structure. But more importantly,
also the angle of incidence has a decisive influence on the symmetry.

\begin{figure}[tb]
\unitlength1cm
\begin{center}
   \begin{picture}(10,7.0)
      \includegraphics{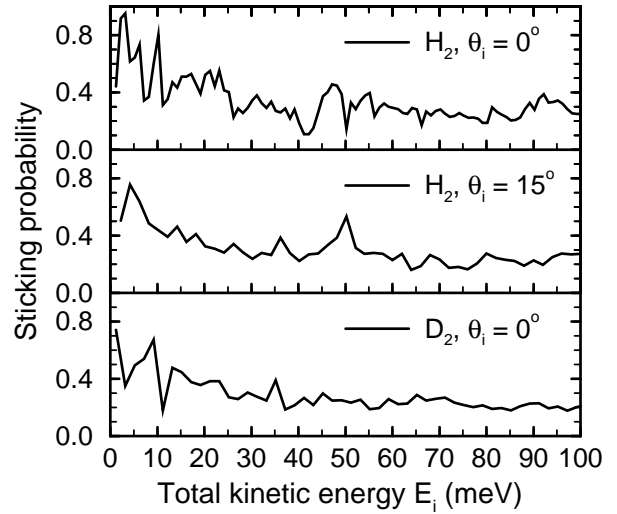}
   \end{picture}

\end{center}
   \caption{5D quantum calculations of the sticking probability 
versus kinetic energy for a hydrogen beam impinging on a Pd(100) surface.
Upper panel: H$_2$ under normal incindence $\theta_i = 0^{\circ}$;
Middle panel: H$_2$ with an incident angle of $\theta_i = 15^{\circ}$;
Lower panel: D$_2$ under normal incidence $\theta_i = 0^{\circ}$.
}

\label{5Dstick}
\end{figure}

The experiment by Rettner and Auerbach was done for an angle of
incidence of 15$^{\circ}$, while the calculations were done
for normal incidence. In order to investigate the dependence of 
the oscillatory structure on the angle of incidence we performed
five-dimensional vibrationally adiabatic calculations for an angle of 
incidence of the molecular beam of 15$^{\circ}$ and compared them
with normal-incidence results (see Fig.~\ref{5Dstick}). 
The energy resolution for the non-normal incidence results
was chosen to be below the width of the prominent peaks for
normal incidence so that these peaks should be detected.
In order to rule out basis set effects, also the results for normal 
incidence in Fig.~\ref{5Dstick} were obtained by five-dimensional 
calculations. 
Caused by the reduced dimensionality, the height of the peaks for normal 
incidence is changed compared to the full six-dimensional calculations.
This indicates the importance of the full dimensionality for the
calculations. The peak positions, however, are the same. Also the 
averaged sticking probability is not changed significantly \cite{Gro96CPLa}.

Now at normal incidence every diffraction
channel is at least fourfold degenerate (except for the backscattered
beam) due to the $C_{4v}$ symmetry of the (100) surface. This makes
the effect of the opening up of a new scattering channel much more
dramatic than in the case of a general angle of incidence, where
this degeneracy is lost. This is demonstrated in Fig.~\ref{5Dstick},
which shows that in the energy regime below
40~meV, which was probed by Rettner and Auerbach,\cite{Ret96PRL,Ret96}
 the sticking probability for $\theta_i = 15^{\circ}$ is
much smoother than for normal incidence. 
Thus, as expected, calculations for normal incidence are not directly
comparable with experiments for non-normal incidence, at least as far as
quantum effects are concerned.

The large peak at approximately $E_i \approx 50$~meV is still visible
for both angles of incidence. This peak is due to the opening
up of rotationally inelastic diffraction, i.e., the kinetic energy
becomes large enough to enable $j = 0 \rightarrow 2$ rotational
transitions in scattering. For $\theta_i = 15^{\circ}$ this peak is
slightly shifted to higher {\em total} kinetic energies. 
This is simply due to the fact that the rotationally inelastic {\em specular}
peak opens up at higher total kinetic energy due to the parallel 
momentum conservation.

In recent quantum dynamical calculations of the dissociative 
adsorption of the reactive system H$_2$/W(100)\cite{Kay95}
also oscillations in the sticking probability were found, 
but they were much smaller than for H$_2$/Pd(100).
This may be caused by the lower dimensionality of these calculations.
Since only one surface coordinate was considered, the number of
degenerate scattering channels opening up was less than in six-dimensional
calculations, leading to smaller effects. For example, in previous quantum
dynamical calculations of the dissociative adsorption of H$_2$
on a model potential with activated as well as non-activated paths
to adsorption, where also only one surface coordinate was considered,
quantum oscillations have also been found,\cite{Gro95JCP} there amplitude,
however, is much smaller than in the 6D-calculations.

For the heavier isotope D$_2$ the energetic spacing between
quantum levels is much less due to the higher mass compared to H$_2$.
The higher ``density'' of channels should also make the effects
of the opening up of new scattering channels less dramatic.
In the lower panel of Fig.~\ref{5Dstick} we have plotted
five-dimensionals results for the D$_2$ sticking probability
under normal incidence with the same energy resolution as for
the H$_2$ non-normal incidence results. Except for low kinetics energies 
the sticking curve is indeed much smoother than for H$_2$ at
normal incidence.

There is a further source for the smoothening of the sticking
probability as a function of the incident kinetic energy in supersonic 
molecular beam experiments: The experimental beam does not correspond
to a monoenergetic beam in one specific quantum state. Instead,
there is a certain velocity spread of the impinging molecules
which is typically of the order of $\Delta v /v_i = 0.1$, where
$v_i$ is the mean initial velocity \cite{Ren89}; in addition,
the internal states of the molecules are populated according to
some Boltzmann-like distribution. For the solid line in 
Fig.~\ref{stickiso} we have assumed an initial rotational and energy
distribution adequate for molecular beam experiments. As a
consequence, the oscillatory structure is almost entirely
washed out. Accordingly, also the experimental data of
Ref.~\onlinecite{Ren89} do not show any significant oscillations.

\subsection{Steering effect}

We will now discuss the general trends in the averaged sticking
probability as a function of the kinetic energy.
The qualitative features of the experimental sticking probability
\cite{Ren89} are well reproduced by the averaged quantum dynamical
results, as Fig.~\ref{stickiso} shows, although there are quantitative
differences which we will address below. At low energies there is 
a substantial decrease in the sticking probability with increasing
kinetic energy, which is then followed by an increase at higher
kinetic energies. As already pointed out above, such a general
behavior had usually been attributed to a precursor mechanism,
in which the impinging molecules prior to dissociation are first
trapped in a physisorption well due to energy transfer to 
substrate phonons \cite{Ren94}. This mechanism, however, cannot
explain the quantum dynamical results since first, there is no
physisorption well in the calculated PES, and second, there is no
energy transfer to the surface possible due to the use
of a fixed substrate. Thus the decrease in the sticking probability
has to be caused by a purely dynamical effect, namely the steering
effect: \cite{Gro95JCP,King78,Gro95a,Wil95,Kay95,Whi96}
Although the majority of pathways to dissociative adsorption
has non-vanishing barriers with a rather broad distribution of
heights and positions, slow molecules can be very efficiently
steered to  non-activated pathways towards dissociative adsorption 
by the attractive forces of the potential. This mechanism becomes 
less effective at higher kinetic energies where the molecules are 
too fast to be focused into favourable configurations towards 
dissociative adsorption.  If the
kinetic energy is further increased, the molecules will eventually
have enough energy to directly traverse the barrier region leading
to the final rise in the sticking probability.

\begin{figure}[tb]
\unitlength1cm
\begin{center}
   \begin{picture}(10,6.5)
      \includegraphics{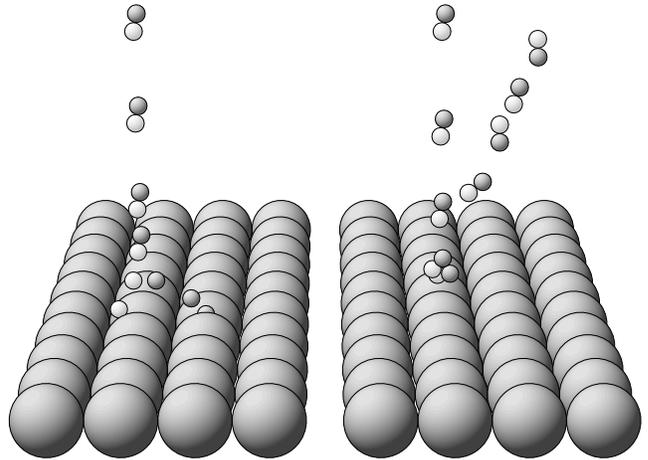}
   \end{picture}

\end{center}
   \caption{Snapshots of classical trajectories of hydrogen molecules 
impinging on a  Pd(100) surface. The initial conditions are chosen in 
such a way that the trajectories are restricted to the $xz$-plane.
Left trajectory: initial kinetic energy $E_i = 0.01$~eV. 
Right trajectory: same initial conditions as in the left trajectory
except that the molecule has a higher kinetic energy of 0.12 eV. }

\label{traj2run}
\end{figure}

In order to illustrate the steering effects, we use the results
of two typical classical trajectory runs.  This is done in 
Fig.~\ref{traj2run}, where snapshots of these two  trajectories
are shown.  The initial conditions are chosen in 
such a way that the trajectories are restricted to the $xz$-plane.
The left trajectory illustrates the steering effect.
The incident kinetic energy is $E_i = 0.01$~eV. 
Initially the molecular axis is almost perpendicular to the
surface. In such a configuration the molecule cannot dissociate
at the surface. But the molecule is so slow that the forces 
acting upon it can reorient the molecule.
It is turned parallel to the surface and then follows a non-activated
path towards dissociative adsorption. 

In the case of the right trajectory, the initial conditions are the same
as in the left one, except  that the molecule has a higher kinetic 
energy of 0.12~eV. Due to the anisotropy of the PES the molecule also
starts to rotate to a configuration parallel to the surface. However,
now the molecule is so fast that it hits the repulsive wall of the potential
before it is in a favorable configuration to dissociative adsorption.
It is then scattered back into the gas-phase rotationally excited.

Fig.~\ref{stickiso} shows that there are still quantitative differences
betweeen theory and experiment. Considering the fact that there are
no adjustable parameters in our calculations, the agreement is quite
satisfactory, though. The discrepancies might be due to uncertainties
in the determination of the {\it ab initio} PES which are of the order
of $\pm 0.1$~eV \cite{Wil95}. We also like to point out that the
experimental values of the sticking probability are subject of
a current debate \cite{Ren89,Ret96}.   

Furthermore, in our calculations substrate phonons or electronic excitations 
are not taken into account. We have noted above that due to the large
mass mismatch between impinging hydrogen molecule and the Pd substrate
atoms the substrate motion can be neglected as far as understanding the
basic dissociation mechanism is concerned. Taking the substrate motion
into account would allow for recoil of the surface atoms upon impact
of the impinging molecules. Although the energy transfer to the
solid is rather small, recoil of the surface atoms leads to a sticking
curve which is stretched to higher energies \cite{Han90,GroErr}. In other
words, it renormalizes the energy axis, because due to the energy transfer
to the surface the effective kinetic energy becomes smaller. Such a 
renormalization would improve the agreement between experiment and
theory, as an inspection of Fig.~\ref{stickiso} reveals.

\subsection{Comparison quantum-classical dynamics and isotope effects}

\begin{figure}[tb]
\unitlength1cm
\begin{center}
   \begin{picture}(10,6.3)
      \includegraphics{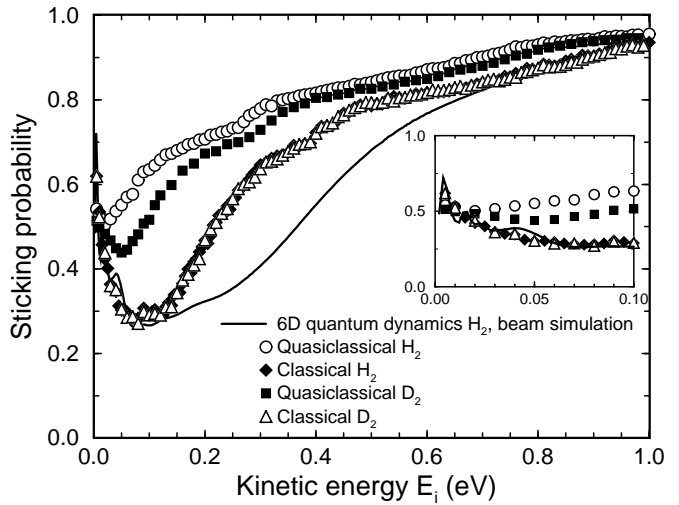}
   \end{picture}

\end{center}
   \caption{Probability for dissociative adsorption
versus kinetic translational energy for a hydrogen beam under normal
incidence on a clean Pd(100) surface for non-rotating molecules.
The solid line shows six-dimensional quantum dynamical 
results for H$_2$ assuming an energy spread typical for beam experiments.
The quasiclassical results corresponds to molecules initially
vibrating with an energy equal to the vibrational zero-point energy
while the classical results are obtained with initially non-vibrating 
molecules. Open diamonds: quasiclassical H$_2$; filled diamonds:
classical H$_2$; filled squares: quasiclassical D$_2$; open
triangles: classical D$_2$.
The inset shows an enlargement of the results at low energies.}

\label{stickclass}
\end{figure}

In Fig.~\ref{stickclass} we compare the averaged quantum mechanical
sticking probability for H$_2$ with the results of classical 
and quasiclassical trajectory calculations for H$_2$ and D$_2$. 
The inset shows an enlargement of the results at low energies.
Quasiclassical in this context corresponds to trajectories
with the initial vibrational energy of the hydrogen molecule
equal to the vibrational zero-point energy of hydrogen, which
is 0.258~eV for H$_2$ and 0.185~eV for D$_2$, while for the purely
classical trajectories the molecules are initially non-vibrating.
First of all, the classical results do not show
any oscillatory structure revealing that the oscillations are entirely
due to quantum mechanics.
Note that the quasiclassical calculations for H$_2$ show almost
no initial decrease in the sticking probability. For D$_2$ there
is a small decrease, while the purely classical results effectively
fall upon the averaged H$_2$ quantum results at low and high
kinetic energies.

We have recently shown that the strong difference between quasiclassical
results on the one side and classical and averaged quantum results
on the other side is caused by zero-point effects of the hydrogen
molecule in the multi-dimensional configuration space \cite{Gro97Vac}. 
When the molecule
approaches the surface, the molecular bond is weakened and consequently
the molecular vibration is softened, i.e., the vibrational frequency
decreases. Since the change of the frequency is slow compared to
the vibrational period, the vibrational energy follows the change of
the frequency almost adiabatically \cite{Gro97Vac,Gro96CPLa} which
leads to an effective vibrational-translational energy transfer.
At the same time, due to the anisotropy and corrugation of
the PES the molecule about to dissociate becomes localized in the
remaining four degrees of freedom of the molecule which are the
polar and azimuthal rotation and the two translations parallel to
the surface. This localization leads to the building up of additional
zero-point energies due to the Heisenberg uncertainty principle. 
In fact, the sum of all zero-point energies
remains approximately constant along a minimum energy 
path towards dissociative adsorption \cite{Ham92,Kra94}, 
and for H$_2$/Pd(100) the sum becomes even 
larger than the gas-phase vibrational zero-point energy
of H$_2$, which is the only zero-point energy of a free molecule.

Now a classical particle can follow precisely a minimum energy path
through a corrugated PES; in  a pictorial sense one might say it
can propagate along the bottom of the valley in the PES. A quantum
particle cannot do that. It has to be delocalized and needs to have at least
the zero-point energies perpendicular to the propagation direction
to traverse a corrugated PES without tunneling. This leads to an effective 
upwards shift of the potential for the quantum particle along the 
minimum energy path. Or {\it vice versa}, the classical particle
experiences a lower minimum potential. As a consequence, the sticking
probability for the quasiclassical particle is much larger than for
the quantum particle.

As mentioned above, the sum of all zero-point energies along the minimum 
energy path towards dissociation adsorption increases in the systen 
H$_2$/Pd(100), but in a first approximation we assume it to be roughly 
constant. In such a situation the combined effect of all zero-point
energies is to cause a constant shift of the potential. Therefore
the results of quantum calculations and classical calculations
without any initial zero-point energy should be similiar since
a constant shift of the potential does not affect the dynamical
properties. This is indeed the case at low and at high energies,
as the comparison between purely classical  and quantum results
in Fig.~\ref{stickclass} demonstrates. In addition, these results confirm
that steering is a general dynamical effect and is not particularly
related to quantum or classical dynamics.

The problem of a proper treatment of zero-point 
energies in quasiclassical trajectory calculations is 
well-known, especially in the gas-phase community \cite{Bow89,Mil89}. 
One possible way to deal with this problem is the reduced 
dimensionality treatment in the vibrationally adiabatic approximation
(for a overview see Ref.~\onlinecite{Cla86}). In this approach a small 
number of degrees of freedom is treated dynamically while the 
remaining degrees of freedom are taken into account by adding the sum of 
their zero-point energies to the potential along the reaction path.
Another more elaborate approach is to constrain the energy in each
vibrational mode to be greater than its zero-point value \cite{Bow89,Mil89}.

In our purely classical approach we ignore zero-point energies
all along the reaction path. But this approach is actually
in the spirit of the vibrationally adiabatic approximation. 
It effectively takes the zero-point energies into account through 
a shift of the potential along the reaction path corresponding to 
the sum of all zero-point energies. This shift, however, is constant
along the reaction path. Moreover, we still keep the full
dimensionality of the problem by explicitly treating all degrees of 
freedom dynamically. This is indeed essential since for example the
steering effect is  absent in a low-dimensional treatment 
of the H$_2$/Pd(100) system \cite{Sch92,Dar92,Bre94}.

Besides zero-point effects tunneling is also an important quantum
phenomenon. However, in a system with activated and non-activated paths
towards dissociative adsorption tunneling does not play an important
role. This is due to the fact that tunneling is exponentially suppressed.
Hence the propagation of the quantum particel along a classically 
possible path is much more probable than the dissociation via tunneling.

The results also show that in purely classical dynamics there is 
no isotope effect between H$_2$ and D$_2$ in the sticking probability.
As far as the low-energy regime is concerned, this seems surprising at
a first glance, since D$_2$ is more inert than H$_2$ due to its higher mass.
However, one has to keep in mind that at the same kinetic energy D$_2$
is slower than H$_2$, so that there is more time for the steering forces 
to redirect the D$_2$ molecule. This has been noted before by 
Kay~{\it et al.}\cite{Kay95}.

Indeed, the Lagrangian for a system of classical particles with the
same mass~$M_1$ can be written as
\begin{equation}
L \ = \ \sum_i \ \frac{M_1}{2} \left( \frac{dx_i}{dt} \right)^2 \ 
- \ U(\{x_i\}) \ . 
\label{Lag_M1}
\end{equation}
For another isotope with the mass~$M_2$ the potential does not change.
If we transform the time axis via
\begin{equation}
t' \ = \ \sqrt{\frac{M_1}{M_2}} \ t \ ,
\end{equation}
we end up with the following Lagrangian for the new isotope of mass~$M_2$:
\begin{equation}
L' \ = \ \sum_i \ \frac{M_1}{2} \left( \frac{dx_i}{dt'} \right)^2 \ 
- \ U(\{x_i\}) \ ,
\label{Lag_M2} 
\end{equation}
which is equivalent to the Lagrangian of Eq.~\ref{Lag_M1}.
This means that the equations of motion for an system of classical particles 
with mass~$M_1$ correspond to the equations of motion for a system of classical
particles with mass~$M_2$, where the velocities have been scaled by an
factor $\sqrt{\frac{M_1}{M_2}}$, i.e., where the kinetic energy is the same.  
Hence, for different isotopes with the same initial conditions,
where only the initial velocities have been scaled to keep
the kinetic energy unchanged, the trajectories remain {\em exactly}
the same.

It follows that there cannot be any isotope effects as a function of the 
kinetic energy for hydrogen moving classically on a PES
as long as there are no energy transfer
processes to, e.g., substrate phonons. {\it Furthermore, this
indicates that the steering effect is not restricted to light molecules
as hydrogen, but should also be operative for all other heavier molecules
moving in a similiar PES.} As far as dissociative adsorption is concerned,
however, for heavier molecules recoil effects of the substrate 
might no longer be negligible, so that for a complete description considering
all relevant degrees of freedom energy transfer processes to
substrate phonons could be important.

Contrary to the purely classical calculations, the quasiclassical results 
show an isotope effect between H$_2$ and D$_2$.
The sticking probability of H$_2$ is larger compared to D$_2$, the effect
being most pronounced for kinetic energies between 0.03~eV and 0.30~eV.
This isotope effect can only be caused by the
different initial vibrational zero-point energies which
can be effectively used to traverse the corrugated and anisotropic
barrier region. The H$_2$ gas-phase zero-point energy is larger by
73~meV; indeed the H$_2$ sticking curve seems to be shifted to lower
energies with regard to the D$_2$ sticking curve by approximately this
amount.

Interestingly enough, at very low kinetic energies below 0.03~eV also
the quasiclassical calculations show almost no isotope effect, in addition
to the fact that classical and quasiclassical results are almost identical
in this low-energy range. In the limit of zero initial kinetic energy
apparently the sticking probability is to a large extent determined
by steering forces which already act rather far away from the surface
where the change of the vibrational frequency and thus zero-point effects
are insignificant. The absence of an isotope effect for very low energies
is actually also true for the averaged quantum results, as 
Figs.~\ref{stickiso} and \ref{5Dstick} demonstrate. However, there is a
pronounced isotope effect in the quantum results for kinetic energies
larger than 0.1~eV. The size of this isotope effect corresponds to the 
one found in the quasiclassical calculations which again shows that the
different initial vibrational zero-point energies cause the isotope 
effect.

We also like to comment on the rather large difference between classical 
and quantum results in Fig.~\ref{stickclass} for kinetic energies
between 0.15~eV and 0.6~eV. 
We think that this difference might be due to the fact
that the sum of all zero-point energies along the minimum energy
path through the barrier region actually becomes larger than the
gas-phase zero-point energy.\cite{Gro97Vac} This effect is most prominent
in the medium energy range, where steering is no longer effective.
At very high kinetic energies, where zero-point effects should play only
a negligible role, indeed quantum and classical results are in very close
agreement. 
Furthermore, in quantum mechanics it always takes
a finite amount of energy to change the state of a particle (if there are
no degenerate states), while in classical mechanics particles can be
diverted by any infinitesimally small amount of energy. This should
make the quantum propagation somehow stiffer than classical propagation.
This could also contribute to the difference between quantum
and classical results in the medium energy range.

\subsection{Non-normal incidence}

\begin{figure}[tb]
\unitlength1cm
\begin{center}
   \begin{picture}(10,6.5)
      \includegraphics{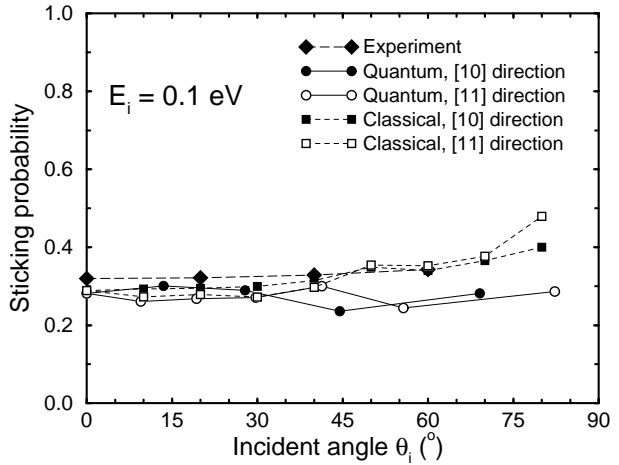}
   \end{picture}

\end{center}
   \caption{Probability for dissociative adsorption of H$_2$ on Pd(100)
versus angle of incidence for an inital total kinetic energy 
of $E_i = 0.1$~eV. Experiment: diamonds 
(from ref.~\protect{\onlinecite{Ren89}}).
The theoretical results are for initially non-rotating molecules. 
Circles show 6D quantum results, squares classical
trajectory calculations. The filled and open symbols correspond to 
calculations with the azimuthal angle of incidence along the [10] and [11] 
direction of the (100) surface, respectively.}
\label{stickangle_0.1}
\end{figure} 

In this  section we address the issue of non-normal incidence.
The experiments on the angular dependence of the sticking probability
of H$_2$/Pd(100) were done for two different initial kinetic energies,
$E_i =0.1$~eV and $E_i = 0.4$~eV.\cite{Ren89} 
The incident azimuth was not identified.
We have determined the quantum and classical angular dependence of the
sticking probability for two different incident azimuths: along
the [10] direction which corresponds to one axis of the surface square
lattice, and along the [11] direction which corresponds to the
diagonal of the surface square lattice.

Fig.~\ref{stickangle_0.1} compares the experiment with the quantum 
and classical results at the lower kinetic energy, $E_i = 0.1$~eV.
Note that the quantum results were determined for a monoenergetic beam
in one specific quantum state which is here the vibrational and
rotational ground state; hence quantum oscillations are superimposed
on these data \cite{Gro95JCP}, but apparently their size is small.
The general experimental trend is well reproduced by the theoretical
results: the sticking probability is almost independent of the 
angle of incidence at this energy, but slightly increases with
increasing angle. There is no large difference between quantum and
classical results. Only at angles larger than 45$^{\circ}$ the classical
results are above the quantum results. There is also almost no
significant dependence on the azimuth except for the classical
results at almost grazing incidence of $\theta_i = 80^{\circ}$ where
the sticking probability along the [11] direction is larger by
0.1 compared to the [10] results.

\begin{figure}[tb]
\unitlength1cm
\begin{center}
   \begin{picture}(10,6.5)
      \includegraphics{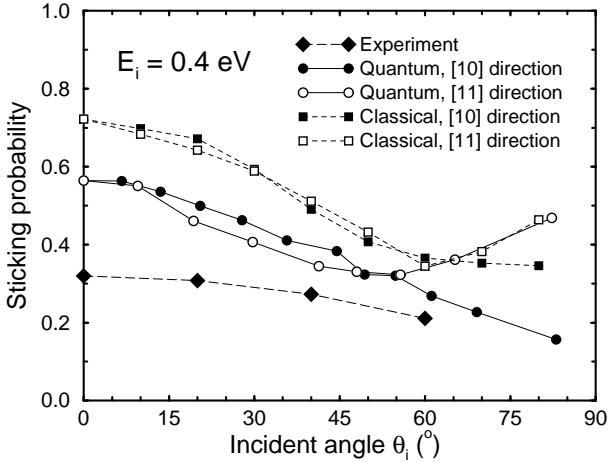}
   \end{picture}

\end{center}
   \caption{Probability for dissociative adsorption of H$_2$ on Pd(100)
versus angle of incidence for an inital total kinetic energy 
of $E_i = 0.4$~eV. The meaning of the symbols is the same as in
Fig.~\protect{\ref{stickangle_0.1}}.
}
\label{stickangle_0.4}
\end{figure}

The angular dependence of the sticking probability at the higher
kinetic energy of $E_i = 0.4$~eV is plotted in Fig.~\ref{stickangle_0.4}.
First of all the absolut values determined by experiment, quantum and
classical calculations are very different at this energy, as is
already apparent from Fig.~\ref{stickiso} and Fig.~\ref{stickclass};
this issue was discussed above. But the general trends in the angular
dependence are in good agreement. All sets of data show 
a significant decrease in the sticking probability with increasing
incident angle for angles below 60$^{\circ}$. For almost grazing
incidence there is now a substantial dependence on the initial azimuth.
The sticking probability for molecules impinging along the [11]~direction
is signifincantly larger than for molecules impinging along the 
[10]~direction. This difference is more pronounced for the quantum than
for the classical calculations.

Our results show that the general features of the angular dependence
of the sticking probability determined in the experiment -- an increase
with increasing incident angle at low energies and a decrease at higher
kinetic energies -- is well reproduced by our six-dimensional calculations.
In particular, the calculations demonstrate that a sticking probability
increasing with increasing incident angle in not necessarily indicative of a 
precursor mechanism \cite{Ren89} but can be caused by the dynamics of the 
dissociative adsorption on a corrugated PES.

Still the questions remains: what causes the different angular dependence
at these two energies?
It is useful to discuss angular effects by considering the energy scaling
of the sticking probability \cite{Dar95}, i.e., by determining the exponent~$n$
such that
\begin{equation}
S (E_i,\theta_i) \approx S (E_i \cos^n \theta_i,\theta_i = 0^{\circ}) 
\label{ergscal}
\end{equation}
If $n =2$, then the so-called normal energy scaling is valid, i.e., the 
sticking probability is a function of the normal component of the incident 
energy alone. In our calculations the sticking probability for normal 
incidence has its minimum at approximately $E_i \approx 0.1$~eV (see 
Fig.~\ref{stickiso}). If normal energy scaling were fulfilled in the 
system H$_2$/Pd(100), then for $E_i \le 0.1$~eV the sticking probability 
would indeed rise with increasing incident angle since the normal energy 
decreases, and it would fall with increasing angle for $E_i > 0.1$~eV as 
long as the normal component $E_i \cos^2 \theta_i$ is larger than 0.1~eV.

\begin{figure}[tb]
\unitlength1cm
\begin{center}
   \begin{picture}(10,6.5)
      \includegraphics{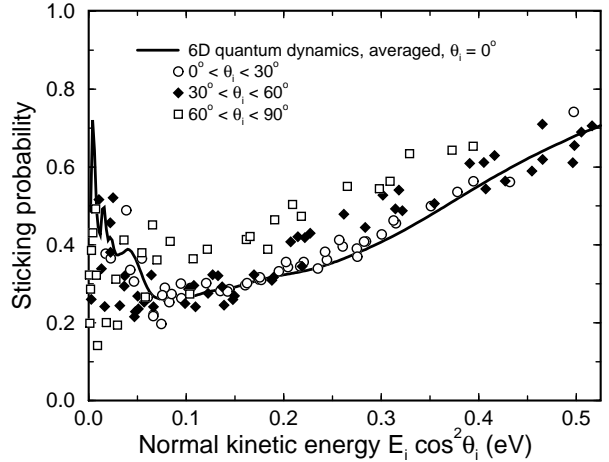}
   \end{picture}

\end{center}
   \caption{Quantum mechanical sticking probability of H$_2$ on Pd(100)
as a function of the normal component of the incident energy
for molecules initially in the vibrational and rotational ground state.
Solid line: beam under normal incidence with an energy spread typical for
molecular beam experiments; circles, diamonds, and squares: monoenergetic
beam with an angle of incidence between 0$^{\circ}$ and 30$^{\circ}$,
between 30$^{\circ}$ and 60$^{\circ}$, and between 60$^{\circ}$ and 
90$^{\circ}$, respectively. The results for non-normal incidence
are obtained for different azimuthal angles.
}
\label{stick_normerg}
\end{figure}

In order to check whether normal energy scaling is valid in the system
H$_2$/Pd(100), we have plotted the quantum mechanical sticking probability
as a function of the normal component of the incident energy 
$E_i \cos^2 \theta_i$ in Fig.~\ref{stick_normerg}. The non-normal 
incidence data show some scatter, in particular at low energies. This 
can be caused by quantum oscillations; the azimuthal dependence, which 
is not specified in Fig.~\ref{stick_normerg}, also contributes to the 
scattering of the data. But the general trends are in qualitative 
agreement with  model calculations on a three-dimensional
PES with activated as well as non-activated path towards dissociative
adsorption \cite{Gro95JCP}. At low normal kinetic energies below 0.05~eV
additional parallel momentum suppresses the sticking, for normal kinetic
energies between 0.05 and 0.35~eV additional parallel momentum enhances
sticking, and above 0.35~eV the results show approximate normal energy
scaling. These results also show that it is not possible to assign any
global energy scaling, i.e. any global exponent $n$ in Eq.~\ref{ergscal},
to the angular dependence of the sticking probability.

At very low energies, where the steering mechanism is operative,
additional parallel momentum hinders dissociation. This can be easily
understood. A molecule with a low normal velocity may still be steered
to a favorable site for dissociation. But due to the additional parallel
momentum the molecule is swept past this favorable site and scattered back 
into the gas-phase from a repulsive site before the bond-breaking can 
occur. This effect is similiar to the rotational hindering in the steering 
regime\cite{Gro95a,Gro96SSb,Beu95,Beu96,Gos97} which is caused by the fact 
that rapidly rotating molecules rotate out of favorable orientation for 
dissociation during the interaction with the surface.
However, Fig.~\ref{stickangle_0.4} shows that for incident angles above
70$^{\circ}$ the suppression of the steering depends strongly on the
incident azimuth.

In the intermediate energy range between 0.05 and 0.35~eV additional 
parallel momentum enhances sticking, in particular for incident angles 
above 60$^{\circ}$. For molecules impinging on the surface under an
angle larger than 60$^{\circ}$ the component of the kinetic energy 
parallel to the surface is at least three times larger than the
normal component. These molecules experience an lateral average of the 
PES in this energy range.\cite{Gro95JCP} Steering in the angular degrees 
of freedom can still occur. Indeed the sticking probability for
$\theta_i > 60^{\circ}$ shows a decrease for normal kinetic energies 
between 0.05~eV and 0.12~eV indicating a steering effect, and then an
increase at higher energies. Far away from the surface the molecules are
first attracted to the on-top-site\cite{Wil95}. But molecules steered
to this site will eventually encounter a barrier towards dissociative
adsorption of 0.15~eV. In order to dissociate slow molecules have to
be re-directed towards the bridge or hollow sites 
(see also Ref.~\onlinecite{Eich96}). Thus potential gradients can also 
steer molecules to ``wrong'' sites. This oversteering in the lateral
coordinates cannot occur for molecules experiencing a laterally
averaged potential causing the increase in the sticking probability  
for large additional parallel momentum. 

In the direct dissociation regime for normal kinetic energies larger 
than the lateral average of the barrier height additional parallel 
momentum causes an increase in the sticking probability.\cite{Gro95JCP}
This lateral average still depends on the orientation of the molecule.
For the majority of molecular orientations the laterally averaged
barrier heights for H$_2$/Pd(100) are less than 0.15~eV 
(see, e.g., the barrier distribution in Ref.~\onlinecite{Gro96CPLa}). 
Hence the lateral averaging also leads to an increase in the 
sticking probability in the direct dissociation regime. 

This mechanism, however, does not promote sticking significantly any
more if the normal kinetic energy is larger than most of the 
maximum barriers for fixed molecular orientation. Still, 
the fact that for $E_i \cos^2 \theta_i >$~0.35~eV the sticking probability 
shows approximate normal energy scaling in spite of the strong corrugation 
of the PES is reminiscent of the activated system H$_2$/Cu. There similiar 
results have been found both experimentally \cite{Ang89,Ret92} and 
theoretically \cite{Gro94} although the PES is also strongly 
corrugated \cite{Ham94,Whi94}. This apparent contradiction is attributed to 
features of the PES called balanced corrugation \cite{Dar95,Dar94SS,Bre95wf}. 
For this type of corrugation the higher barriers have to be farther away 
from the surface compared to the lower barriers. These features are also 
present in the system H$_2$/Pd(100) where the highest barriers are over 
the on-top-sites \cite{Wil95}.

We now return to the discussion of the angular dependence of the
sticking probability for fixed total kinetic energy. For fixed
total kinetic energy increasing the incident angle means decreasing
the normal kinetic energy and increasing the incident parallel momentum.
At low kinetic energies decreasing the normal kinetic energy makes
the steering more effective which promotes dissociation. On the other
hand, increasing the incident parallel momentum hinders dissociation
in the low-energy range. At $E_i = 0.1$~eV both effects approximately
cancel which leads to a sticking probability almost independent of the
incident angle (see Fig.~\ref{stickangle_0.1}).  

At normal energies larger than 0.1~eV decreasing the normal kinetic 
energy leads to a decrease in the sticking probability, but increasing 
the incident angle enhances the sticking probability for normal energies
below 0.4~eV. However, the promoting effect of additional parallel momentum 
is less pronounced than the decrease due to the smaller normal kinetic
energy. Hence in Fig.~\ref{stickangle_0.4} the sticking probability
decreases for increasing incident angle at an initial total kinetic
energy of 0.4~eV.

Figs.~\ref{stickangle_0.1} and \ref{stickangle_0.4} also show that for 
$\theta_i < 60^{\circ}$ there is almost no dependence of the sticking 
probability for non-normal incidence on the azimuth. For larger incident
angles, however, molecules impinging along the [11] direction of
the surface unit cell have a higher dissociation probability than
molecules impinging along the [10] direction. This can be explained
by a shadowing effect. For molecules approaching the surface under an almost
grazing incidence along one axis of the quadratic surface unit cell, the most 
favorable adsorption path at the bridge position is effectively hidden
behind the high barriers at the on-top position. For an approach along
the diagonal of the square unit cell this most favorable adsorption path is
still directly accessible.

\begin{figure}[tb]
\unitlength1cm
\begin{center}
   \begin{picture}(10,6.5)
      \includegraphics{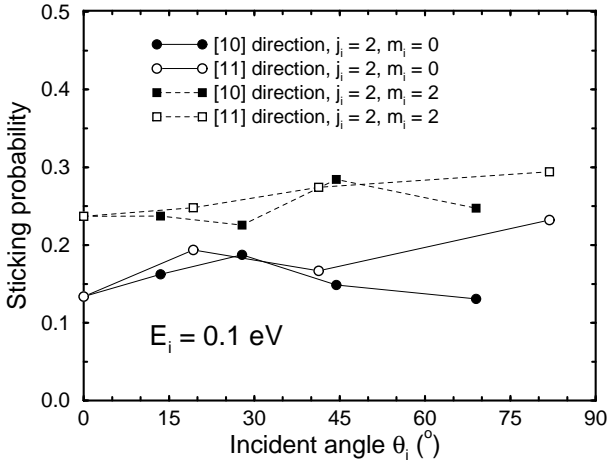}
   \end{picture}

\end{center}
   \caption{Quantum dynamical probability for dissociative adsorption 
of H$_2$ on Pd(100) versus angle of incidence for an inital 
total kinetic energy of $E_i = 0.1$~eV.
The results are for molecules initially in the rotational quantum
state $j = 2$. 
Circles show results for initial rotational azimuthal quantum number
$m_i = 0$, squares for $m_i = 2$. The filled and open symbols correspond to 
calculations with the azimuthal angle of incidence along the [10] and [11] 
direction of the (100) surface, respectively.
}
\label{stickangle_0.1_l2}
\end{figure}

Finally we study the influence of the incident rotational quantum
state on the angular dependence of the sticking probability.
We have determined the sticking probability as a function
of the angle of incidence for molecules initially in the
rotational quantum state $j_i =2, m_i =0$ and $j_i =2, m_i =2$ for
$E_i = 0.1$~eV. The results are plotted in Fig.~\ref{stickangle_0.1_l2}.
Note that for $\theta_i = 0^{\circ}$ the sticking probability for 
initially non-rotating molecules with $j_i =0$ is 0.3 
(see Fig.~\ref{stickangle_0.1}). Fig.~\ref{stickangle_0.1_l2} shows
the well-known result\cite{Gro95a,Gro96SSb,Beu95,Beu96,Gos97} that rotational 
motion hinders the dissociation at low energies because rotating molecules
rotate out of favorable orientations for dissociation. This suppression, 
however, depends on the orientation of the molecules. Molecules with azimuthal
quantum number~$m = j$ have their axis preferentially oriented parallel to the 
surface. These molecules rotating in the so-called helicopter fashion  
dissociate more easily than molecules rotating in the cartwheel fashion 
($m = 0$) where the rotational axis is preferentially oriented parallel 
to the surface. The latter have a high probability hitting the
surface in an upright orientation in which they cannot dissociate. 

This steric effect is effective for all incident angles, i.e., the
sticking probability for $m_i = 2$ is always larger than for $m_i = 0$.
Like for the non-rotating molecules at this kinetic energy, the results 
show only a weak dependence on the incident angle.
For incident angles below 45$^{\circ}$ there is also no significant
dependence on the azimuth, but again, for almost grazing incidence
molecules approaching along the diagonal of the surface unit
cell have a higher dissociation probability than molecules 
approaching along one axis of the surface unit cell.
These results indicate that to first order rotational and parallel
motion are decoupled as far as the dissociation dynamics is concerned.

\section{Conclusions}

In conclusion, we reported  a six-dimensional quantum and classical 
dynamical study of dissociative adsorption of hydrogen on Pd(100).
We used a potential energy surface obtained from detailed density 
functional theory calculations for the system H$_2$/Pd(100). 
The six hydrogen degrees of freedom are treated fully dynamically.
The two main approximations are, firstly, that the substrate is kept
fixed so that no thermal disorder or phonon excitations are allowed,
and secondly, that the system is assumed to remain in its electronic
ground state. Hence the continuous excitation spectrum of the semi-infinite 
substrate is neglected. Still these calculations are able to reproduce 
all of the known experimental results with regard to the 
disscociative adsorption at least semi-quantitatively. 
The time-reverse process to dissociative adsorption, 
the associative desorption, was not discussed in this study,
but previous studies showed that also experimental desorption properties
are well-reproduced by our calculations.
Among the processes that are now quite well understood are the
dependence of adsorption and/or desorption on the molecular translational 
energy, vibrational and rotational state, orientation of the molecule,
and the angle of incidence.

Quantum effects are non-neglible for hydrogen dissociation on surfaces. 
The discrete nature of diffraction and molecular excitation leads to
a strong oscillatory structure of the sticking probability as a 
function of the incident kinetic energy. Furthermore, 
zero-point effects  cause substantial deviations between 
{\em averaged} quantum dynamical calculations and
quasiclassical calculations, in which the initial conditions
correspond to a molecule vibrating with the gas-phase zero-point
energy of hydrogen. The corrugated and anisotropic potential energy 
surface leads to the building up of additional zero-point energies 
which effectively increase the minimum potential in the quantum  
calculations. This changes the dynamics in the low-energy regime
of H$_2$/Pd(100) dramatically. However, the building up of the
additional zero-point energies roughly cancels and even overcompensates
the decrease in the zero-point energy of the H-H vibrations upon dissociative
adsorption. Therefore purely classical calculations which neglect
the zero-point energies in the initial conditions are closer to the
quantum results than the  quasiclassical calculations.

At low kinetic energies the dissociative adsorption is dominated by the
steering effect. For higher kinetic energies steering becomes less
efficient leading to the initial decrease in the sticking probability.
The steering effect is dependent on the kinetic energy, but not on
the mass of the molecule. Hence steering should also be effective
for heavier molecules.

There are still some quantitative differences between theory and
experiment. They might be caused by uncertainties in the evaluation
of the PES, but also by uncertainties in the experimental determination
of the sticking probability. In addition, the differences might be
caused by the neglect of substrate phonons or electronic excitations
in the calculations. Hence we will address the role of the substrate 
degrees of freedom in the adsorption and desorption processes in the
future. As for now, our results show that the {\it ab initio} determination 
of the potential energy surface combined with high-dimensional dynamical 
calculations, in which the relevant degrees of freedon are taken into account, 
is an important step forward in our understanding of simple reactions at 
surfaces.

\end{document}